\documentclass[final,5p,times]{elsarticle}
\usepackage{booktabs}
\usepackage{tabularx}

\usepackage{hyperref}

\usepackage{pgf-umlcd}
\usepackage{comment}

\usepackage{multirow}

\usepackage{amsmath}
\usepackage{amssymb}

\usepackage{color}

\usepackage{algorithm}
\usepackage{algpseudocode}

\usepackage{multicol}

\usepackage{amsmath}
\usepackage{graphicx}
\usepackage{xcolor}
\usepackage{soul}
\usepackage[normalem]{ulem}

\graphicspath{ {./} }

\usepackage{caption}
\definecolor{cadetblue}{rgb}{0.37, 0.62, 0.63}
\definecolor{fuzzywuzzy}{rgb}{0.8, 0.4, 0.4}
\definecolor{babyblue}{rgb}{0.54, 0.81, 0.94}

\usepackage{tikz}
\usetikzlibrary{calc, matrix, positioning}
\usetikzlibrary{shapes, arrows, arrows.meta, shadows, positioning}

\tikzset{
  frame/.style={
    rectangle, draw,
    text centered,
    text width=6em, 
    minimum height=4em,fill=white,
    rounded corners,
  },
  initialize/.style={
    rectangle, draw,
    text centered,
    text width=16em, 
    minimum height=4em,fill=white,
    rounded corners,
  },
  block/.style={
    draw, rectangle, 
    text centered,
    fill=white,
    minimum height=3em, 
    minimum width=5em,
    rounded corners,
  },
  sum/.style={
    draw, circle,
    fill=white,
    node distance=1cm,
  },
  input/.style={coordinate},
  output/.style={coordinate},
  line/.style={
    draw, -{Latex},rounded corners=3mm,
  },
  pinstyle/.style={
    pin edge={to-,thin,black}
  }
}

\usepackage{subcaption} 

\usepackage[noabbrev]{cleveref}

\usepackage{amssymb}
\journal{Engineering Applications of Artificial Intelligence}

\begin{document}

\begin{frontmatter}

\title{Deep Learning Based Simulators for the Phosphorus Removal Process Control in Wastewater Treatment via Deep Reinforcement Learning Algorithms}

\author[inst1,inst2]{Esmaeel Mohammadi\corref{cor1}}
\ead{esm@kruger.dk}
\cortext[cor1]{Corresponding author.}
\author[inst1]{Mikkel Stokholm-Bjerregaard}
\author[inst1]{Aviaja Anna Hansen}
\author[inst2]{Per Halkjær Nielsen}
\author[inst3]{Daniel Ortiz-Arroyo}
\author[inst3]{Petar Durdevic}
\ead{pdl@energy.aau.dk}

\affiliation[inst1]{organization={Krüger A/S},%Department and Organization
            addressline={Indkildevej 6C}, 
            city={Aalborg},
            postcode={9210}, 
            state={North Jutland},
            country={Denmark}}

\affiliation[inst2]{organization={Department of Chemistry and Bioscience, Aalborg University},%Department and Organization
            addressline={Fredrik Bajers Vej 7H}, 
            city={Aalborg},
            postcode={9220}, 
            state={North Jutland},
            country={Denmark}}

\affiliation[inst3]{organization={AAU Energy, Aalborg University},%Department and Organization
            addressline={Niels Bohrs vej 8}, 
            city={Esbjerg},
            postcode={6700}, 
            state={South Jutland},
            country={Denmark}}

\begin{abstract}
Phosphorus removal is vital in wastewater treatment to reduce reliance on limited resources. Deep reinforcement learning (DRL) is a machine learning technique that can optimize complex and nonlinear systems, including the processes in wastewater treatment plants, by learning control policies through trial and error. However, applying DRL to chemical and biological processes is challenging due to the need for accurate simulators. This study trained six models to identify the phosphorus removal process and used them to create a simulator for the DRL environment. Although the models achieved high accuracy ($>$97\%), uncertainty and incorrect prediction behavior limited their performance as simulators over longer horizons. Compounding errors in the models' predictions were identified as one of the causes of this problem. This approach for improving process control involves creating simulation environments for DRL algorithms, using data from supervisory control and data acquisition (SCADA) systems with a sufficient historical horizon without complex system modeling or parameter estimation.
\end{abstract}

% \begin{minipage}{\textwidth}
% \begin{highlights}
% \item The proposed method facilitates the training and % implementation of deep reinforcement learning algorithms for biological systems
% \item Investigates the state-of-the-art models' potential for creating deep reinforcement learning environment simulators
% \item States the challenges of creating simulation environments for highly dynamic processes
% \end{highlights}
% \end{minipage}

\begin{keyword}
Deep Reinforcement Learning \sep Dynamic model \sep Simulator \sep Attention \sep Sequence to sequence \sep Phosphorus
% %% PACS codes here, in the form: \PACS code \sep code
% %% MSC codes here, in the form: \MSC code \sep code
% %% or \MSC[2008] code \sep code (2000 is the default)
\end{keyword}
\end{frontmatter}


\section{Introduction}
Phosphorus (P) is essential for human nutrition and plant growth. Without it, the primary cells of plants, animals, and life would not exist \cite{Nobaharan2021, Melgaco2021}. The amount of available P is limited due to the decreasing number of phosphate rock resources \cite{porter2009phosphorus}. The phosphorus nutrient is often found in wastewater treatment plants (WWTP) outlets in surface waters; therefore, high P concentrations lead to eutrophication that can affect the environment and human health \cite{welch1980ecological}. Efficient phosphorus removal from wastewater prevents eutrophication and its consequences \cite{GU2021}. It is possible to remove phosphorus from wastewater by incorporating phosphate into Total Suspended Solids (TSS) and subsequently removing it from them. A P-containing bio-solid (microorganisms, for example) or chemical precipitate can be formed \cite{tchobanoglous2003wastewater}. Removal and recovery of P can be promoted by analyzing its dynamics through the wastewater treatment process, which helps engineers and operators comprehensively understand what’s happening and identify potential future problems \cite{HANSEN2022}. The dynamic control system measures and controls various parameters of the wastewater treatment process, such as pH, dissolved oxygen, temperature, and flow rate. With real-time monitoring of these parameters, operators can optimize treatment efficiency, reduce energy consumption, and minimize waste production and byproduct generation \cite{Revollar2017}. The enhanced Biological phosphorus Removal (EBPR) process, which takes advantage of phosphorus accumulating organisms (PAO), is a more environmental-friendly and cost-effective way of phosphorus treatment rather than chemical removal methods which use the addition of metal salts for P precipitation \cite{acevedo2014modelling}. Combining chemical and biological processes can achieve a desirable level of P removal with lower expenses. Therefore, most plants nowadays use a hybrid P removal system \cite{bunce2018review}.

Control of the phosphorus removal systems in wastewater treatment plants is challenging in current literature because, on the one hand, variables like dissolved oxygen (DO) are not the best indicators of the aerobic zone. On the other hand, there are uncertainties in incoming phosphate concentration and the influence of process conditions (pH, temperature, etc.) on the coagulation process \cite{chong2013assessing, seviour2003microbiology}. Although there is not a wide range of studies on P removal control strategies in the literature due to the mentioned issues, some researchers applied control methods like fuzzy control \cite{xu2015application}, Model Predictive Control (MPC) \cite{OSTACE2013164}, and the Supervisory and Override Control Approach (SOPCA) \cite{SHEIK2022132346} in the framework of benchmark simulation models (BSM) developed by International Water Association (IWA) \cite{alex2008benchmark, henze2006activated}. Benchmark simulation models are standard generalized tools for the simulation of biological processes in WWTPs. Still, they are also not a good representative of the process dynamics, especially P removal, which is highly stochastic and unpredictable. Novel control strategies such as Deep Reinforcement Learning (DRL) algorithms have been recently introduced to overcome the complexity, uncertainties, and challenges of process control for different biological systems \cite{CHEN2021130498}.

Reinforcement Learning (RL) is a machine learning technique that involves training an agent to make decisions based on rewards received from an environment. The agent learns to maximize rewards by taking actions that lead to positive outcomes and avoiding actions that lead to adverse outcomes. Reinforcement learning has been successfully applied in various applications, including game playing \cite{silver2017mastering, vinyals2019grandmaster}, robotics \cite{hua2021learning}, and control systems \cite{sutton2018reinforcement, moriyama2018reinforcement}. Deep reinforcement learning is a variation of reinforcement learning that uses deep neural networks to represent the agent's policy and value functions \cite{sutton2018reinforcement}. DRL can learn from raw sensory data, such as images, audio, or text, and handle high-dimensional inputs and outputs \cite{mnih2015human}. Traditional optimal control methods, such as model predictive control (MPC), face limitations when applied to large-scale stochastic multiple-input multiple-output (MIMO) problems due to their online computational requirements and assumptions about uncertainty information \cite{maravelias2009integration}. On the other hand, DRL can pre-compute optimal solutions offline, reducing online computation time, and can be trained in a process simulator to acquire a general knowledge of the process \cite{NIAN2020106886}. Although the DRL agent's performance may not surpass that of a corresponding MPC designed based on the same simulator model, DRL's learned optimal policy implicitly includes information about optimal set points and inputs, akin to the concept of economic MPC \cite{NIAN2020106886}. As a result, DRL shows promise in the process control industry and has been used considerably in recent control research \cite{Bao2021, Lillicrap2015, moriyama2018reinforcement, raju2015reinforcement, spielberg2017deep}.

In \cite{NIAN2020106886}, some of the shortcomings of reinforcement learning have been discussed for process control applications. In summary, it can be data inefficiency, scalability, stability, convergence, constraints, and accurate simulator. The lack of a precise simulator for most industrial process control applications has been the main issue for using RL methods \cite{NIAN2020106886}. This problem also extends to chemical and biological processes, which led researchers to try implementing different strategies for creating a simulation environment to train and test RL algorithms in it \cite{spielberg2017deep, wang2018novel, nian2019fault}. The application of DRL methods in wastewater treatment plants, especially for phosphorus removal processes, is very limited. In \cite{PANG2019893}, a Q-learning algorithm was used to optimize aerobic and anaerobic hydraulic retention time (HRT) for the biological phosphorus removal process. They used the ASM2d model to generate state transition matrices to train the Q-learning algorithm. The developed model was verified using data from a lab-scale sequencing batch reactor (SBR) with aerobic and anaerobic processes. A Multi-Agent Deep Deterministic Policy Gradient (MADDPG) was used in \cite{CHEN2021130498} for the purpose of dissolved oxygen and chemical dosage control in a WWTP. They used the MANTIS model, an integrated form of three other mechanistic models \cite{CHEN2021130498} to simulate the target WWTP. 

Modeling the wastewater treatment process to study the system behavior has been going on for decades, from earlier mechanistic models \cite{Mogens2000} to advanced data-driven approaches \cite{HANSEN2022, NEWHART2019}. Simple wastewater treatment models such as the Activated Sludge Model (ASM) and Anaerobic Digestion Model (ADM) have been used for years to study the dynamics of WWTPs \cite{Gujer2006,processes2002anaerobic}. These models are no longer feasible and accurate enough to describe different processes in wastewater treatment plants \cite{Burton2014}. Activated Sludge Models No.2 are mainly used for biological phosphorus removal modeling \cite{Mogens2000}. Anaerobic Digestion Model No.1 (ADM1) has also been used for studying struvite precipitation \cite{ikumi2011} and biodegradability of organics in anaerobic digestion \cite{ikumi2014}.

With the advance of Artificial Intelligence (AI), modeling of WWTPs based on machine learning and deep learning methods has become very popular \cite{YE2020}. AI methods can predict operational parameters, evaluate energy usage, fault diagnosis, automation, and intelligent control in WWTPs \cite{Malviya2021, ZHAO2020}. Data-driven modeling of WWTPs has emerged as an alternative to mechanistic models, as the former does not necessitate a thorough comprehension of the plant's design and operation. Furthermore, data-driven models can be developed relatively swiftly and with fewer input data, as per the findings of \cite{NEWHART2019}. Artificial Neural Networks (ANN), including the Multi-layer Perceptron (MLP) network, have been demonstrated as a robust and precise technique for forecasting operational parameters in WWTPs \cite{MANNINA2019, nelles2020nonlinear, wunsch2018forecasting, pisa2019artificial}. Data from WWTPs can be treated as time series. The auto-regressive integrated moving average (ARIMA) model is a time series prediction method that uses past data to predict future target values \cite{BERTHOUEX1996}. ARIMA has been used for studies like sedimentation modeling \cite{park2015} and water quality prediction \cite{OMERFARUK2010}. A Recurrent Neural Network (RNN), an extended feed-forward Neural Network (FFNN), can be used for time series prediction. RNN can capture and pass information through its elements with the help of memory and takes advantage of past information for decision-making \cite{CHENG2019, ZHU2020}. RNNs have been used in controlling activated sludge process \cite{foscoliano2016}, forecasting the water flow of the WWTP \cite{zhang2018manage}, predicting the amount of ammonium and total nitrogen \cite{pisa2019ann}, as a software sensor for prediction of BOD, COD, and TSS indexes \cite{CHANG2021}, and as a fuzzy controller for the dissolved oxygen and nitrate concentration in WWTP \cite{Gaitang2016}.

As a type of recurrent neural network, Long Short-term Memory (\emph{LSTM}) networks can learn order dependence for sequence prediction \cite{sak2014}. The two technical problems of conventional RNNs overcome by \emph{LSTM}s are vanishing and exploding gradients related to how the network is trained \cite{Graves2009}. Considering the good results of implementing \emph{LSTM}s in areas like speech recognition \cite{PENG2021} and natural language processing \cite{SHUANG2020}, it has been used in wastewater treatment studies recently. In the wastewater treatment literature, \emph{LSTM}s have been implemented for different kinds of applications such as prediction of effluent quality \cite{Pisa2019}, forecasting wastewater flow rate \cite{Kang2020}, estimation of ammonium, total nitrogen, and total phosphorus removal efficiency \cite{YAQUB2020}, predicting influent BOD, effluent BOD, temperature, and power efficiency \cite{CHENG2020}, and as a control strategy in WWTPs \cite{Pisa2020Control}. In \cite{HANSEN2022}, an \emph{LSTM} model was implemented to predict phosphorus dynamics in wastewater treatment plants. They used Bayesian optimization for hyperparameter tuning of the model, which could predict phosphorus concentrations up to 24 hours in the future.

As a result of Vaswani et al. 's \emph{Transformer}, encoder-decoder models based on attention mechanism were generated \cite{vaswani2017attention}. \emph{Transformer}s have been used to forecast time series in environmental engineering areas such as defect detection inside the sewage system \cite{DANG2022} and prediction of effluent water quality in WWTPs \cite{Huang2021}. Later, \emph{Informer} \cite{zhou2021informer} was introduced in 2021 to overcome some problems of the vanilla \emph{Transformer}, such as high memory usage and limitations of the encoder-decoder architecture for time-series forecasting. \emph{Informer} could outperform existing time-series forecasting methods like ARIMA, \emph{LSTM}, and \emph{Transformer}s for the benchmark datasets of energy and weather \cite{zhou2021informer}. \emph{Autoformer} is a recent version of \emph{Transformer}s introduced in \cite{wu2021Autoformer} to perform long-time series forecasting. Currently, no study is based on the \emph{Informer} and \emph{Autoformer} models for wastewater treatment applications. Still, considering that the input sequence length for wastewater parameters is uncertain, whether these approaches will perform well in WWTP modeling is a question. The abovementioned deep learning models can be used to simulate the DRL agent's environment. The model inputs the current state of the environment and outputs the next state and the reward signal for the agent.

The lack of an accurate simulator for implementing deep reinforcement learning algorithms in WWTPs, motivated us to study state-of-the-art deep learning models to address this problem. For this purpose, we first extracted a dataset containing information on the P removal process from the target WWTP (Agtrup, Denmark) and formulated it as a time-series prediction problem. Then we trained six models to use them as simulation environments to implement DRL algorithms. These models had different architectures for time series prediction, such as linear (\textbf{\emph{LTSF Linear}} \cite{Zeng2022AreTE}), recurrent neural networks (\textbf{\emph{LSTM}} \cite{Sepp1997LongSM}), attention-based (\textbf{\emph{Transformer}} \cite{vaswani2017attention}, \textbf{\emph{Informer}} \cite{zhou2021informer}), and auto-correlation (\textbf{\emph{Autoformer}} \cite{wu2021Autoformer}). We intended to use the various models to explore different approaches to identify the most effective time series forecasting model for implementing deep reinforcement learning algorithms. Through this exploration, we aimed to understand the challenges inherent in modeling such biological processes and identify the best path forward for designing an accurate simulator. The differences in the models' structure influenced the simulation environment results, where we found the strengths and weaknesses of each prediction mechanism.

\section{Plant}
This study focuses on Kolding central WWTP in Agtrup Denmark, which has a 125,000 population equivalents (PE) capacity and a current load of approximately 65.5\% \cite{Agtrup2021}. The plant removes phosphorus with a combination of chemical and biological removal methods. In the chemical phosphorus removal, metal salts such as aluminum sulfate (alum), ferrous sulfate, or ferric chloride are added to the wastewater in a rapid mix tank, followed by flocculation to form a precipitate with soluble phosphorus \cite{Burton2014}. On the other hand, biological phosphorus removal relies on naturally occurring microorganisms called phosphorus Accumulating Organisms (PAO). PAOs release stored phosphorus under anaerobic conditions and remove soluble phosphorus under aerobic conditions \cite{ZHANG2022118102}. Both processes can be challenging to control, as inlet phosphate concentrations can vary unpredictably due to industrial contributions. Conversion of polyphosphate to orthophosphate before coagulant addition will affect coagulation efficiency in chemical removal. Additionally, cultivating a population of organisms that can survive alternating anaerobic and aerobic cycles requires long acclimation periods from a few days to several weeks or even months \cite{Tuszynska2019}. Moreover, high recycle nitrate concentrations can inhibit anaerobic zone processes for biological removal \cite{chong2013assessing, seviour2003microbiology}.

\subsection{Operation}
\label{operation}
\begin{figure}
\includegraphics[width=\columnwidth,,height=\textheight,keepaspectratio]{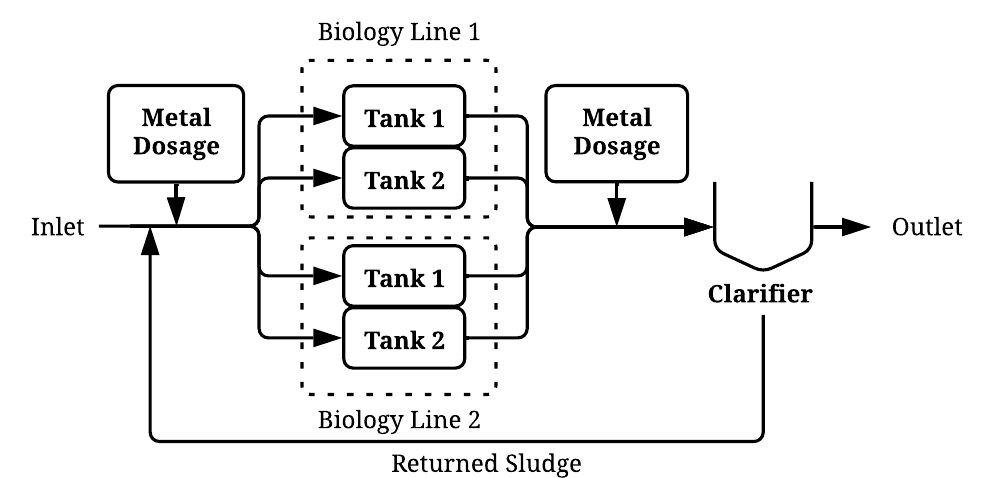}
\centering
\caption{Schematic of the phosphorus removal process in the plant with the flow lines}
\label{fig:agtrup_diagram}
\end{figure}

The current operation of the plant for phosphorus removal is shown in Figure \ref{fig:agtrup_diagram}. The system consists of a wastewater treatment plant where iron salts are added at two locations: after the primary settler and before the secondary settler. The biological phosphorus removal part is placed between the two settlers and consists of 2 parallel lines, each including two reaction tanks. Only tank 1 in biology line 1 has the phosphate sensor. The plant is currently being monitored and controlled by the Hubgrade$^{TM}$ Performance Plant system, designed by Krüger/Veolia. The sampling frequency is specified to measure the plant's outputs and the controller's updates, which is currently every two minutes.

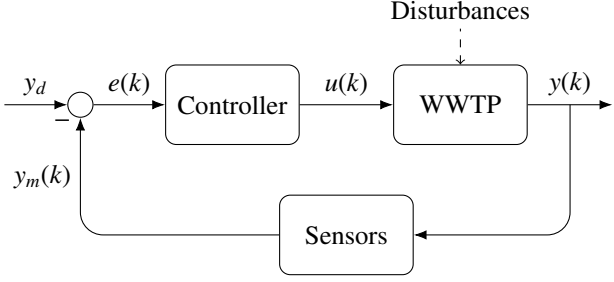
\begin{figure}
\centering
\begin{tikzpicture}[auto, node distance=2cm,>=latex']
    \node [input, name=input] {};
    \node [sum, right of=input] (sum) {};
    \node [block, right of=sum] (controller) {Controller};
    \node [block, right of=controller, pin={[pinstyle]above:Disturbances},
            node distance=3cm] (system) {WWTP};

    \draw [line] (controller) -- node[name=u] {$u(k)$} (system);
    \node [output, right of=system] (output) {};
    \node [block, below of=u] (measurements) {Sensors};

    \draw [draw,line] (input) -- node {$y_d$} (sum);
    \draw [line] (sum) -- node {$e(k)$} (controller);
    \draw [line] (system) -- node [name=y] {$y(k)$}(output);
    \draw [line] (y) |- (measurements);
    \draw [line] (measurements) -| node[pos=0.99] {$-$} 
        node [near end] {$y_m(k)$} (sum);
\end{tikzpicture}
\caption{The control block diagram of the system}
\label{fig:control_block_diagram}
\end{figure}

The control strategy shown in Figure \ref{fig:control_block_diagram} involves a feedback control algorithm that calculates the manipulated variable (metal dosage) based on the measured outputs and the setpoint. The specific algorithm used and any limitations or assumptions of the strategy are described in detail in \cite{HANSEN2022}. In general, the control strategy can be described by the following equation:
\begin{equation}
 u(k) = K(y_d - y_m(k))   
\end{equation}

Where $u(k)$ is the manipulated variable at time $k$, $K$ is the controller gain, $y_d$ is the setpoint or reference value for the phosphate concentration in the outlet, and $y_m(k)$ is the measured concentration at time $k$. The system responds to changes in the setpoint or disturbances, with limitations on its performance. The iron salts are dosed at specified locations and timing using a dosing mechanism based on the commands from the explained controller.

\subsection{Dataset Preprocessing}
\label{dataset}
Nineteen months of data ($D$) were collected from the SCADA system at the Agtrup plant between June 2021 and January 2023. $D$ consisted of 23 two-dimensional vectors, each representing a pair of variables at the treatment plant. Specifically, D can be represented as a matrix of size $n \times m$, where $n$ is the number of measurements taken over the 19 months, and $m$ is the number of variables measured. Each row of the matrix corresponds to a single measurement, and each column corresponds to a particular variable. The value of each variable is represented by the first component of the corresponding vector, while the second component represents the quality of the measurement. The quality variable takes binary values, with 0 and 1 indicating good and bad quality, respectively.

After preprocessing the data, to detect issues such as insufficient quality data, negative values, and missing values, the dataset with 23 variables has been processed with feature selection or engineering methods. The Pearson correlation method in the \emph{pandas} library \cite{reback2020pandas} was used to investigate the most critical variables of the plant affecting the phosphate concentration. The selected variables were: Nitrate concentration [mg/L], Ammonia concentration [mg/L], and Ammonia plus Nitrate concentration [mg/L], which are added to the metal dosage flow [m\textsuperscript{3}/hr] and Phosphate concentration [mg/L] to form the final dataset. Additionally, Gradient Boosting Regression with decision trees was used to determine the most optimum way of using time features (hour, day of the week, and month) in the dataset to improve the prediction accuracy.

\section{Models}
\label{models}
The following will explain the types of different deep learning models used in the simulation environment for the phosphorus removal process in wastewater treatment plants. All models were built with the \emph{PyTorch} library \cite{NEURIPS2019_9015}.

\subsection{Long Short-Term Memory (\emph{LSTM})}
The LSTM was introduced by \cite{Sepp1997LongSM} as a successful technique for addressing the vanishing gradient problem in recurrent neural networks. Unlike conventional RNNs, which use a recurrent node, the LSTM structure replaces the recurrent node with a memory cell. The memory cell contains a self-connected recurrent edge of fixed weight 1, which allows the gradient to pass through many time steps without disappearing or exploding. The memory cell is also equipped with multiple multiplicative gates, such as the input gate, forget gate and output gate. These gates determine whether to impact the internal state of the neuron, flush the internal state to zero, or allow the internal state to influence the cell's output, respectively. \emph{LSTM} models have become popular for time series forecasting and system identification due to their ability to capture long-term dependencies in the data and handle variable-length input sequences \cite{HANSEN2022, Wang2017}. The power of \emph{LSTM} to fit non-linear long periodic data patterns like wastewater treatment data has resulted in its increased application in this field \cite{Yunpeng2017}. Despite the power of \emph{LSTM} in time series forecasting, it sometimes has trouble capturing long-term dependencies, for example, in language modeling \cite{khandelwal2018sharp}, and the model cannot be parallelized.

\subsection{Transformer}
An important part of the \emph{Transformer} model is the attention mechanism, originally intended to be a sequence-to-sequence RNN improvement; it was used to enhance encoder-decoder RNNs for machine translation \cite{bahdanau2014neural}. Instead of compressing the input, attention suggests that the decoder revisit the input sequence at every step rather than compressing it. The decoder might be able to focus on particular parts of the input sequence at particular decoding steps rather than always seeing the exact representation of the input. As each step of the decoding process was performed, the attention mechanism enabled the decoder to dynamically attend to a different part of the input \cite{bahdanau2014neural}. The \emph{Transformer} architecture for machine translation was proposed by \cite{vaswani2017attention}, which dispenses with recurrent connections and incorporates cleverly arranged attention mechanisms instead of recurrent connections. As a result of its outstanding performance, the \emph{Transformer} began to appear in most of the advanced natural language processing systems by 2018. The attention mechanism of the \emph{Transformer} can be described by the following equation \cite{vaswani2017attention}:
\begin{equation}
Attention(Q,K,V) = softmax(\frac{QK^T}{\sqrt{d_k}})V
\label{attention}
\end{equation}

Here, $Q$, $K$, and $V$ represent the queries, keys, and values matrices. Furthermore, $d_k$ represents the dimension of queries and keys. The softmax function converts the vector of numbers to the vector of probabilities.

The most critical layer of the \emph{Transformer} is self-Attention which means compression of attention toward itself. Compared to previous recurrent and convolutional architectures, self-Attention has the following advantages: parallel operations (compared to RNN) and no need for deep networks to find long sentences \cite{shaw2018self}. As self-attention can lead to loss of ordering information and the \emph{Transformer}, the encoder has no recurrence like RNNs; the model takes advantage of positional encoding to preserve information about the order of tokens \cite{shaw2018self, vaswani2017attention}. Compared to the \emph{LSTM}, the \emph{Transformer} is potentially more successful in capturing recurrent patterns with long-term dependencies. The model can access any part of history, no matter the distance \cite{lai2018modeling}. However, using the \emph{Transformer} for extremely long sequences requires an immense computation power because the growth of space complexity in self-attention is quadratic \cite{huang2018music}, which can lead to problems in forecasting data with substantial long-term dependencies \cite{lai2018modeling}. Also, the \emph{Transformer} uses autoregression to decode the output, resulting in error accumulation in long-term predictions \cite{Zeng2022AreTE}.

\subsection{Informer}
Applying the \emph{Transformer} model to long-term time series forecasting (LTSF) has critical issues such as quadratic time complexity, high memory usage, and the inherent limitation of the encoder-decoder architecture \cite{zhou2021informer}. The \emph{Informer} model was introduced in \cite{zhou2021informer} to overcome the limitations of the \emph{Transformer}-based model for LTSF applications. To do so, \emph{Informer} introduced some new features:  
\begin{itemize}
    \item Implementing the ProbSparse self-attention mechanism is proposed to reduce canonical self-attention. It achieves the complexity and memory usage of $O(L\log L)$, where $L$ is the number of layers in the network.
    \item Attention scores are dominated by self-attention distilling operations.
    \item Improvements in prediction accuracy in LTSF problems, which contain \emph{Transformer}-like models for capturing individual time-series dependency.
    \item Reduced space complexity to $O((2 - \epsilon)L\log L)$, where the parameter $\epsilon$ is a small positive constant that controls the method's accuracy.
    \item A generative style decoder was introduced to obtain long sequences with only one forward step.
\end{itemize}

The \emph{Informer} employs a generative decoder with direct multi-step (DMS) forecasting, which can produce more accurate predictions as it doesn't use the last prediction results as input for the next step \cite{zhou2021informer, Zeng2022AreTE}. 

\subsection{Autoformer}
Using residuals and encoder decoders, \emph{Autoformer} reconstructs the \emph{Transformer} into a decomposition forecasting architecture using residuals and encoder decoders \cite{wu2021Autoformer}. \emph{Autoformer} replaces attention with the Auto-Correlation mechanism, which calculates the relationship between the current value of the variable and its past values. Auto-correlation explores period-based dependencies by counting the series autocorrelations and aggregating similar subseries by time delay.

\subsection{LTSF Linear}
Autoregressive or iterated multi-step (IMS) prediction of the long-term time series data has been proven to have some problems, like the significant accumulation of errors \cite{Zeng2022AreTE}. In \cite{Zeng2022AreTE}, authors challenge the effectiveness of \emph{Transformer}-based models for long-term time series prediction by direct multi-step (DMS) forecasting methods, which have better results when the prediction horizon is large. They argue that applying self-attention to time series data can lead to loss of ordering information, and positional encoding of the data cannot help preserve the temporal information. They claim that the time series data must have significant trends and seasonality to make a long-term prediction. Calling them embarrassingly simple models, they introduce a set of LTSF Linear baselines to compare with \emph{Transformer}-based methods. LTSF Linear outperforms \emph{Transformer}-based models up to 20 - 50\% in long-term time series forecasting, and they prove that in contrast to the past claims, the performance of \emph{Transformer}s will not be improved by increasing the look-back window size \cite{Zeng2022AreTE}. 

LTSF Linear is the simplest direct multi-step model with a linear temporal layer. The basic linear model uses a weighted sum operation of historical time series to predict the future. Considering the weighted sum operation, we will have the \linebreak following:
\begin{equation}
\mathbf{\hat{X}_i} = \mathbf{W} \times \mathbf{X_i}
\label{DLinear_main}     
\end{equation}

In the above formula, $\mathbf{\hat{X}_i}$ is the prediction, and $\mathbf{X_i}$ is the input for each variable (i). The linear layer is shown by $\mathbf{W}$ where $\mathbf{W} \in \mathbb{R}^{T \times L}$.

\emph{DLinear} is one type of LTSF Linear model with a specific data preprocessing method. They used the decomposition layer introduced in previous \emph{Autoformer} \cite{wu2021Autoformer} and FEDformer \cite{zhou2022fedformer}, then combined it with linear layers. It extracts two different components from the raw data, one is a trend component, and the other one is a reminder or a seasonal component. It gets the final prediction by summing up the results from two linear layers applied to each component \cite{Zeng2022AreTE}.
\emph{NLinear} is another linear model specifically used when a distribution shift is found in the dataset. Firstly, \emph{NLinear} subtracts the input by the last value of the sequence as a simple normalization and then sends the input through a linear layer. At the end of the procedure, the subtracted part is added to the output when making the final prediction \cite{Zeng2022AreTE}.

\subsection{Training of the Models}
Each one of the models was trained individually with the {\emph{PyTorch} library on the dataset from the actual plant as described in section \ref{dataset} as input. The dataset was divided between training and testing parts with a ratio of 85\%/15\%, respectively, and 15\% of the training dataset was used for validation. The input to the model at each training step was a sequence of time steps with length $l$, including all the features of the WWTP dataset. The output for each step was the values of all the features for one time step after the last step in the input.

At each time step, $\mathbf{x} = (x_1, \dots , x_n)$ denotes the values of the system's features. The sequence length ($l$) was set at 240, representing 240 minutes of the historical data as input. To ensure all models are trained in the best way possible, \emph{Optuna} \cite{optuna_2019} library is used for hyper-parameter optimization. Optuna's Tree-structured Parzen Estimator (TPE) uses a sampler derived from Bayesian \cite{mockus2012bayesian} optimization. Using TPE, Optuna finds points closer to previous good results rather than at random \cite{optuna_2019}. After this step, the models were trained with the optimized hyper-parameters by using \emph{Adam} as the optimization method and mean squared errors (\emph{mse}) as the loss function. The script from \cite{Zeng2022AreTE}, which is available on \href{https://github.com/cure-lab/LTSF-Linear}{GitHub} was used with some changes to train DLinear and NLinear models. The best checkpoint for each model, which had the lowest \emph{mse} for the validation dataset, was saved for further use in the simulation environment.

\section{Formulation of the Simulator}
Let us consider a nonlinear dynamical system with a vector of manipulated variables ($\mathbf{u}$), a vector of measured outputs ($\mathbf{y}$), a vector of unmeasured disturbances ($\mathbf{w}$), and a vector of state variables ($\mathbf{x}$). The following set of equations can represent this system \cite{brunton_kutz_2019, astrom2013computer, Wang2017}:

\textbf{State} equation:
\begin{equation}
\mathbf{x}(k+1) = \mathbf{f} (k,x(k),u(k),w(k)) 
\label{state_equation}
\end{equation}

Where $\mathbf{x}(k)$, $\mathbf{u}(k)$, and $\mathbf{w}(k)$ represent the state variables, the manipulated variables, and the unmeasured input disturbances at time $k$ respectively.

\textbf{Output} equation:
\begin{equation}
\mathbf{y}(k) = \mathbf{g} (k,x(k),u(k),v(k))
\end{equation}  

Where $\mathbf{y}(k)$, and $\mathbf{v}(k)$ represent the measured output and the measured disturbances at time $k$. In the above equations, $\mathbf{f}$ represents the system's dynamics, while $\mathbf{g}$ represents the output equation. The function f describes how the state variable changes over time, given the values of the manipulated variable, measured output, and unmeasured disturbance. The function $\mathbf{g}$, on the other hand, describes how the state variable influences the measured output. The representation of the unmeasured disturbances in the system dynamics model needs disturbance models for both input and output. It requires accurate disturbance data, which is not possible for all systems. Here, we assume that the unmeasured disturbances ($\mathbf{w}(k)$) are included in the state variables data ($\mathbf{x}(k)$) and result in an estimation of the true states for every step. 

Now, let's consider a dynamical system with input history length $l$ and manipulated variables $\mathbf{u}(k)$ and output at time $k+1$ given by \cite{electronics11182935, Zar2021}:
\begin{equation}
\begin{split}
{\mathbf{x}(k+1)} = \mathbf{f}(x(k),x(k-1),...,x(k-l+1),\\
u(k),u(k-1),...,u(k-l+1))
\end{split}
\label{dynamic_history_x_k}
\end{equation}

In the above equation, $f$ is a nonlinear function that describes how the system evolves over time based on the input history $\mathbf{x}(k), \mathbf{x}(k-1), ..., \mathbf{x}(k-l+1)$, and the manipulated variables $\mathbf{u}(k), \mathbf{u}(k-1), ..., \mathbf{u}(k-l+1)$. In the current study, we use six deep learning models described in section \ref{models} as the $f$ function stated in equation \ref{dynamic_history_x_k}.

Here, we will explain the \emph{LSTM} as an example of how a deep learning model can describe a specific discrete-time dynamical system. The basic \emph{LSTM} model consists of three layers: the input layer, the \emph{LSTM} layer, and the fully connected layer \cite{Sepp1997LongSM}. In order to achieve the value of $\mathbf{x}(k+1)$ in equation \ref{dynamic_history_x_k}, the state variables $(\mathbf{x}(k-l),...,\mathbf{x}(k))$, and the manipulated variables $(\mathbf{u}(k-l),...,\mathbf{u}(k))$ first will be passed through the \emph{LSTM} layer. In this layer, the cell input activation vector ($\tilde{\mathbf{c}}$), the cell state vector ($\mathbf{c}$), and the hidden state vector ($\mathbf{h}$) will be computed at each time point $t \in [k-l,\;k]$ by the following equations \cite{Sepp1997LongSM}:
\begin{equation}
{\tilde{\mathbf{c}}}(t) = \sigma_{c}(\mathbf{W}_{c}(\mathbf{x}(t),\mathbf{u}(t)) + \ \mathbf{U}_{c}\mathbf{h}(t-1) + \mathbf{b}_{c})
\label{LSTM_1}
\end{equation}
\begin{equation}
\mathbf{c}(t) = \mathbf{f}(t) \cdot \mathbf{c}(t-1) + \mathbf{i}(t) \cdot {\widetilde{\mathbf{c}}}(t)
\label{LSTM_2}
\end{equation}
\begin{equation}
\mathbf{h}(t) = \mathbf{o}(t) \cdot \sigma_{h}(\mathbf{c}(t))
\label{LSTM_3}
\end{equation}

Where $\mathbf{W}_c$, $\mathbf{U}_c$, and $\mathbf{b}_c$ are the weight matrix, recursive weights matrix, and the bias components of the \emph{LSTM} cell, respectively, additionally, $\mathbf{f}(t)$, $\mathbf{i}(t)$, and $\mathbf{o}(t)$ represent the forget, input, and output gates activation vectors. Also, $\sigma_{c}$ and $\sigma_{h}$ are the sigmoid and the hyperbolic tangent functions. After this step, the output of the \emph{LSTM} layer, which is its hidden state ($\mathbf{h}$), will be passed through a fully connected layer. According to \cite{Sepp1997LongSM, Zar2021, electronics11182935}, the output from the \emph{LSTM} model for a dynamic system can be computed as follows:
\begin{equation}
\mathbf{x}(k+1) = \mathbf{W}_x\mathbf{h}(k) + \mathbf{b}_x
\label{LSTM_fc}
\end{equation}

$\mathbf{W}_x$ and $\mathbf{b}_x$ are the weight vector and the bias components of the fully connected layer at the output of the \emph{LSTM} model. $\mathbf{h}(k)$ is the hidden state or the output vector of the \emph{LSTM} layer at time $k$. The final output of the \emph{LSTM} network is the system's state variables $\mathbf{x}$ at time $k+1$.

The explained method can be used to identify and simulate dynamic systems. This identification approach can also be utilized as a train and test environment for process control with deep reinforcement learning research in different industries. To do so, the first step will be formulating the industrial process as a discrete-time dynamical system. Then, the system function or the model should be specified and fit the historical data of the process. Once the trained model is ready, we can create the specific environment with the saved model and used dataset.

Generally, a reinforcement learning environment returns the agent four variables at each step. These four variables are the current or predicted state of the system, reward, a Boolean value specifying whether the episode has been finished or not called done, and info that consists of different information such as the current step of the simulation, total reward, and history of observations. Total reward, or cumulative reward, refers to the sum of rewards obtained by an agent over a sequence of interactions with an environment. It quantifies the performance of the agent in achieving its goals. In continuous tasks, the cumulative reward is often discounted by a gamma ($\gamma$) factor at each time step which is calculated over $T$ steps as \cite{Lillicrap2015}: 
\begin{equation}
R_t = \displaystyle\sum_{i=t}^{T} \gamma^{(i-t)}\mathbf{r}(\mathbf{s_i},a_i)
\label{comulative_reward}
\end{equation}

In the above equation, $R_t$ is the cumulative reward, $\mathbf{r}$ is the reward function where $\mathbf{s_i}$ and $a_i$ are the system's state and the action taken in the step $i$. The agent will work based on the returned variables from the environment and generate the action ($a$), which in our case, is the flow of metal dosage in the process. The action from the agent will replace the current metal addition in the state vector. The predicted state will be added to the state's history, and the process will be repeated. The whole procedure is shown in Figure \ref{fig:env_test_diagram}, which will be further explained in the following.
\begin{figure}
\centering
\begin{tikzpicture}[auto, node distance = 4cm]
    \node [frame] (agent) {Agent};
    \node [frame, below=1.2cm of agent] (environment) {Simulation Environment};
    
    \draw[line] (agent) -- ++ (3.5,0) |- (environment)
    node[right,pos=0.25,align=left] {action\\ $a_t$};
    \coordinate[left=8mm of environment] (P);
    \draw[thin,dashed] (P|-environment.north) -- (P|-environment.south);
    \draw[line] (environment.200) -- (P |- environment.200)
    node[midway,above]{$\hat{s}_{t+1}$};
    \draw[line,thick] (environment.160) -- (P |- environment.160)
    node[midway,above]{$r_{t+1}$};
    \draw[line] (P |- environment.200) -- ++ (-1.4,0) |- (agent.160)
    node[left, pos=0.25, align=right] {state\\ $s_t$};
    \draw[line,thick] (P |- environment.160) -- ++ (-0.8,0) |- (agent.200)
    node[right,pos=0.25,align=left] {reward\\ $R_t$};
\end{tikzpicture}
\caption{The process of a testing simulation environment for one model in a specific time point}
\label{fig:env_test_diagram}
\end{figure}
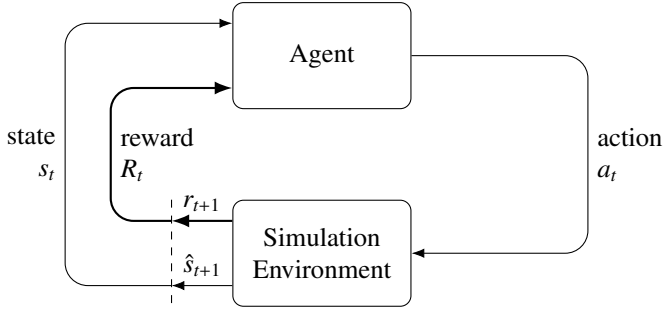

We created a test script to study the accuracy of the simulation environment with the six different models. The script takes one model as the predictor and starts prediction with a series of data from the actual dataset. We give each test step the episode length for which the environment must run. The algorithm for testing the developed environment is shown in algorithm \ref{env_test_code}.

First, the trained model and dataset from the real plant are loaded, as explained in section \ref{dataset}. Then, some parameters are defined, such as: 
\begin{itemize}
    \item \textbf{episode length}: the number of episodes (time steps) we want to test our simulation environment
    \item \textbf{$i_a$}: the index of action column in the dataset
    \item \textbf{$i_o$}: the index of the objective variable column in the dataset
    \item \textbf{p}: the date-time index of the initial input to the simulation environment
    \item \textbf{N}: a parameter for specifying the simulation step
    \item \textbf{l}: sequence length of the historical input data used in the model training
    \item \textbf{n}: number of the state variables (features) in the dataset
\end{itemize} 

In deep reinforcement learning, the state of the system at each time step $t$ is defined as $\mathbf{s_t}$, and when using a model as $\mathbf{\hat{s}}_t$ because it's a prediction. We can rewrite equation \ref{state_equation} as below for a DRL simulation environment:
\begin{equation}
{\mathbf{\hat{s}}_{t+1}} = f(\mathbf{s_t},a_t)
\label{DRL_state}
\end{equation}

Where $\mathbf{\hat{s}}_{t+1}$ is the predicted state of the system at time $t+1$, and $a_t$ is the action taken from the agent at time $t$. The input to the environment ($\mathbf{s_0}$) is initialized from the time point \textbf{p} with the size of $l \times n$. This input is sent to the environment as the system's current state, where the next state will be calculated according to equation \ref{DRL_state}.

Then, the predicted state of the system ($\mathbf{\hat{s}}_{t+1}$) is delivered to the test agent to produce an action. The agent produces an action ($a_{t+1}$), the real amount of the action variable collected from the plant. Finally, the action ($a_{t+1}$) is added to the input data ($\mathbf{s_{t+1}}$) for the next state and replaces the predicted amount of action variable. This process is repeated for the number of \textbf{episodes} specified, and the values of action, predicted objective variable, and real objective variable from the dataset are saved for each time step.

\begin{algorithm}
\caption{DRL simulation environment testing}\label{Env_Test}
\label{env_test_code}
\begin{algorithmic}
 \State \textbf{Inputs:}
  \Statex \hspace*{\algorithmicindent} \textit{dataset, model, p, M (episode length), and l}
 \State \textbf{Outputs:}
  \Statex \hspace*{\algorithmicindent} \textit{The Simulated States over M episodes}
 \State Initialize:
  \Statex \hspace*{\algorithmicindent} $t \gets 0$
  \Statex \hspace*{\algorithmicindent} $s_0$ from the point \textit{p} of the dataset with size l $\times$ n
  \Statex \hspace*{\algorithmicindent} $a_0$ is the control variable from the last time step in $s_0$
 \For{episode = 1, M}
    \State $\hat{s}_{t+1} \gets \textit{model}(s_t,a_t)$
    \State append $\hat{s}_{t+1}$ to $s_t$ and update 
    \State append $\hat{s}_{t+1}$ to \textit{Simulated States}
    \State update $a_{t+1}$ from the dataset 
    \State $t \gets t+1$
 \EndFor
\end{algorithmic}
\end{algorithm}

By doing so, we are able to compare the real and simulated phosphorus amount at each step and observe the accuracy of the simulation environment in case of having a real-time controller. The real phosphorus amount comes from the original dataset, and the simulated one is extracted from the state of the environment.

\section{Results}
This section will present the results from the models' training and testing of the simulation environment for all of them. The results are discussed in section \ref{discussion}.

\subsection{Models}
After training the models in section \ref{models}, the prediction results for the test dataset were compared. Table \ref{hyper-parameter} shows optimized hyper-parameters for all models using Bayesian optimization. The test dataset's prediction results are shown in Figure \ref{fig:test_results} for all models. The $mse$ reported in Figure \ref{fig:test_results} is the mean squared error of the model's prediction for the test dataset.

\begin{table}
\caption{Optimized hyper-parameters for all models}
\label{hyper-parameter}
\resizebox{\columnwidth}{!}{%
\begin{tabular}{@{}lccccc@{}}
\toprule
\multicolumn{1}{c}{Models} & \multicolumn{5}{c}{Hyper-parameters}        \\ \midrule
 &
  \multicolumn{1}{l}{Learning Rate} &
  \multicolumn{1}{l}{Dropout} &
  \multicolumn{1}{l}{Batch-size} &
  \multicolumn{1}{l}{Number of Layers} &
  \multicolumn{1}{l}{Dimension of Layers} \\
\emph{LSTM}                       & 1e-6 & 0.1 & 32 & 2                   & 249 \\
\emph{Transformer}                & 1e-7 & 0.1 & 64 & 2 (Enc.) - 1 (Dec.) & 512 \\
\emph{Informer}                   & 1e-7 & 0.1 & 64 & 2 (Enc.) - 2 (Dec.) & 512 \\
\emph{Autoformer}                 & 1e-7 & 0.1 & 64 & 2 (Enc.) - 1 (Dec.) & 512 \\
\emph{DLinear}                    & 1e-6 & 0.1 & 16 & -                   & -   \\
\emph{NLinear}                    & 1e-6 & 0.1 & 16 & -                   & -   \\ \bottomrule
\end{tabular}%
}
\end{table}

% This figure should be printed in color.
\begin{figure*}
\includegraphics[scale=1]{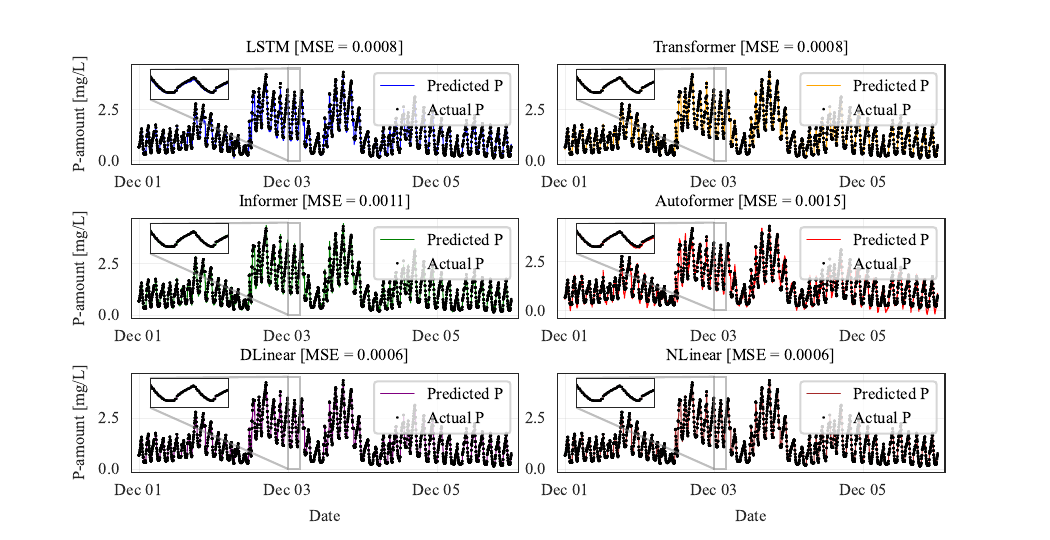}
\centering
\caption{Prediction results of all models for a part (5 days) of the test dataset}
\label{fig:test_results}
\end{figure*}

\subsection{Simulation Environment}
The developed simulation environment was tested for all the models explained in section \ref{models}. In order to get a comprehensive overview of the simulator results with the models, four sequences were chosen from different seasons. The four sequences were selected to explore the behavior of the models during different periods of the year when the data varied in terms of trends, disturbances, and sensor failures. The results of testing the simulation environment for different sequences are shown in Figure \ref{fig:env_results_season}. It is worth noting that the selection of these sequences was based on the characteristics of the data and did not relate to any seasonal studies.
% This figure should be printed in color.
\begin{figure*}
\includegraphics[width=\textwidth,height=\textheight,keepaspectratio]{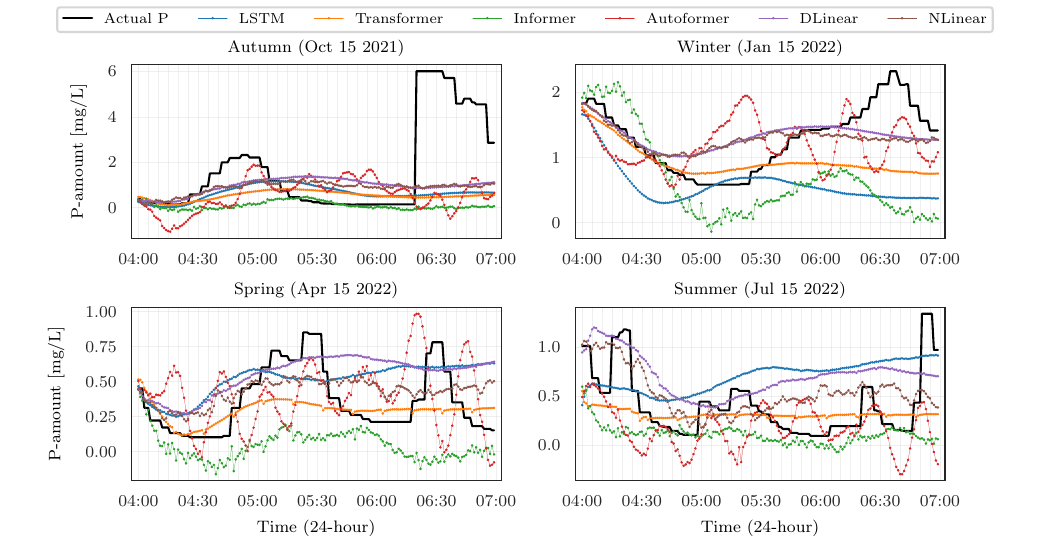}
\centering
\caption{Simulation environment results of all models for the four data points in different sequences}
\label{fig:env_results_season}
\end{figure*}

\section{Discussion}
\label{discussion}
Developing accurate simulation environments for phosphorus removal systems is a big challenge in DRL research due to the complexity and non-linearity of such systems in wastewater treatment plants. The results from this study show that even the state-of-the-art models in the field of time series forecasting might face some problems regarding biological process simulation. Figure \ref{fig:env_results_season} shows the difficulties of deep learning models in predicting the phosphorus removal system's behavior. As one may notice, there is not a specific model that significantly outperforms others, and each one of them has more accurate or the more poor predictions for specific periods.

The possible main reason for inaccuracy and uncertainty in the models' predicted states is an issue called \emph{compounding error} in reinforcement learning \cite{xiao2019}. The compounding error refers to the increase of unacceptable one-step prediction errors over longer horizons \cite{xiao2019, asadi2019, lambert2022}. According to \cite{lambert2022}, the state of a dynamical system after $h$ episodes can be predicted by the following equation:
\begin{equation}
{\mathbf{\hat{s}}_{t+h}} = f(...f(f(\mathbf{s}_t,a_t),a_{t+1})...,a_{t+h})
\label{states_normal}     
\end{equation}

The prediction error for each episode can be defined as $e_t = \mathbf{\hat{s}}_t - \mathbf{s}_t$, increasing multiplicatively because each step's input consists of previously predicted states. This issue can be formulated like the following over a prediction horizon (episode length) of $h$ \cite{lambert2022}:
\begin{equation}
{\mathbf{\hat{s}}_{t+h}} = f(...f(f(\mathbf{s}_t,a_t)+e_t,a_{t+1})+e_{t+1}...,a_{t+h})+e_{t+h}
\label{states_error}     
\end{equation}

Then, the prediction error for each time step can be computed as the mean squared error over all of the state dimensions:
\begin{equation}
MSE = \displaystyle\sum_{d=0}^{d_s} (\mathbf{\hat{s}}_{t,d}-\mathbf{s}_{t,d})^2
\end{equation}

In the above equation, $d_s$ indicates the last index of the state dimensions, which for our problem is $n-1$, and $n$ indicates the number of features. Figure \ref{fig:mse_changes} shows the effect of compounding error in each simulation step for all models in different seasons. The figure shows fluctuations in the mean squared error ($mse$) across different models, with a notable increase in its value observed at later stages compared to earlier stages. The normalized $mse$ values presented in some plots range from 0, representing the most accurate predictions, to 1, indicating completely erroneous predictions. These results suggest that the models may produce unrealistic predictions during extended episodes, and caution should be exercised in interpreting long-term simulations.
% This figure should be printed in color.
\begin{figure*}
\includegraphics[width=\textwidth,height=\textheight,keepaspectratio]{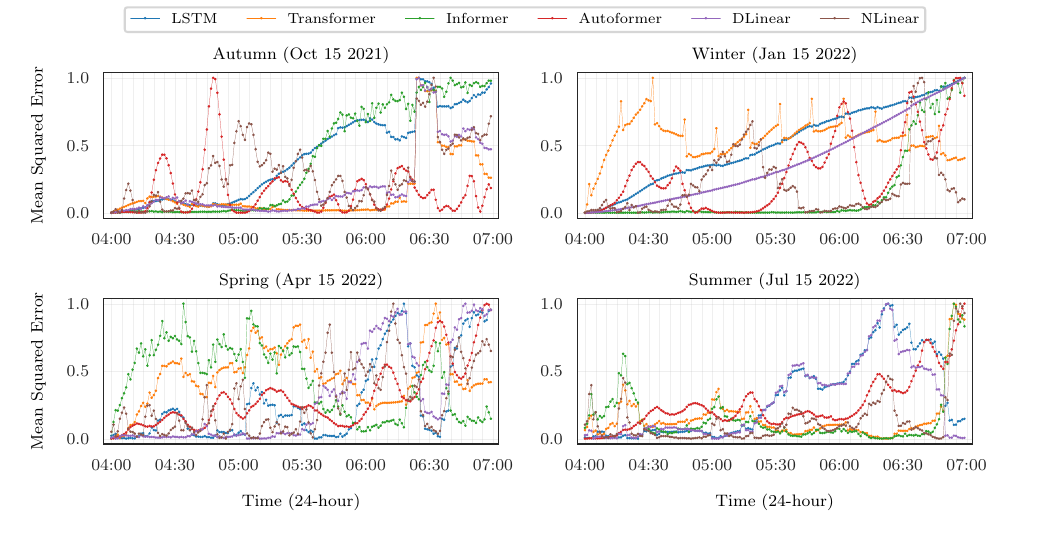}
\centering
\caption{Step-wise mean squared error of all models for the four data points in different sequences}
\label{fig:mse_changes}
\end{figure*}

According to the previous descriptions, we can conclude that the compounding error issue can lead the system model toward unrealistic and uncertain predictions. This issue appears in different forms for each model, which we will discuss in the following sections.

\subsection{LSTM}
The \emph{LSTM} model can capture the system's dynamics in the early steps, but as simulation goes on, it moves towards linear and constant values. Considering the issue of compounding error, the \emph{LSTM} model could avoid extremely fluctuating predictions due to its ability to extract the information from the previous data as possible and capture long-term dependencies. In the summer sequence of Figure \ref{fig:env_results_season}, the \emph{LSTM} has very close predictions to linear models, which results in the inability to capture the system's dynamics.

\subsection{Formers}
The former models (\emph{Transformer}, \emph{Informer}, and \emph{Autoformer}) are different in behavior compared to the \emph{LSTM}. They act better in following the system dynamics but poorly predict the values. In the \emph{Transformer}, autoregression in the decoder results in error accumulation in long-term predictions \cite{Zeng2022AreTE}. As explained in the previous section, the compounding error is the main challenge in simulating the phosphorus removal system's function. The formers also suffer from the compounding error issue, but more different than \emph{LSTM}. These models extract much information from the previous data at each step, leading them to extract many more wrong values. They learn to pay attention to real data information, which results in better predictions. Still, it will be more troublesome when this attention is applied to the predicted and non-precise data. They suffer from \textit{too much} attention. This particular issue is notably evident in the spring sequence shown in Figure \ref{fig:env_results_season}, specifically regarding the \emph{Autoformer} model. In this instance, the model generates excessively dynamic predictions surpassing actual values.

\subsection{LTSF Linear}
\emph{DLinear} and \emph{NLinear} have more similar behavior to the \emph{LSTM}, as they can keep the system dynamics at the beginning of the simulation. Still, they lean towards a linear constant prediction as it goes forward. On the one hand, the more simple and linear models cannot extract much information from historical data, resulting in the inability to capture system dynamics. On the other hand, their simple structure prevents them from forecasting far from the actual values, leading to a less fluctuating simulation. According to Figure \ref{fig:env_results_season}, this behavior can be seen in all sequences but is more obvious in the autumn and winter parts.

\subsection{Common Issues}
As the simulation behavior differs, the models share common issues in the results. The first one is that none of them captures the sudden changes in the process, which is more evident at the end of the autumn sequence (from 06:20 to 06:35) and in the summer sequence (from 06:50 to 07:00) in Figure \ref{fig:env_results_season}. These changes can be related to sensor failures and inaccurate data from the actual plant, which is unknown, and there is no possibility of including them in the models. Even though the models were trained on the dataset with such inaccurate data, they were successful in not following the wrong dynamics.

One potential issue arises in the summer sequence, where the initial prediction differs from the actual observation. This disparity highlights the possibility of models generating slightly wrong predictions at the onset of the simulation, particularly when the entire input sequence is derived from the actual dataset. Consequently, models that begin with less precise predictions may exhibit inferior performance in subsequent episodes compared to those that initially generated accurate predictions.

While the models may generate similar values that differ slightly in dynamics, there are instances where the predicted values can significantly diverge. Such a scenario is apparent in the winter sequence of Figure \ref{fig:env_results_season}, indicating that the behavior of models relative to one another can vary depending on the input sequence.

\section{Conclusion}
The lack of accurate simulators presents a significant challenge in implementing Deep Reinforcement Learning for chemical and biological processes. In this study, the authors trained six different state-of-the-art time series forecasting models to create a simulator for the DRL environment. These models showed more than 97\% accuracy in test dataset prediction. However, the performance of these models was limited by uncertainty and incorrect prediction behavior over longer time horizons. While the \emph{LSTM} and \emph{Linear} models moved towards linear predictions over time and couldn't capture the system's dynamics, the \emph{Former}, models had more fluctuating behavior as the simulation continued. In conclusion, using time series models as a simulation environment where the predicted state at each step is used as input to the next step leads to problems such as compounding errors. 

To address this challenge, further research is needed to develop simulation environments that don't suffer from the compounding error problem and can be implemented for the chemical and biological processes. Future research can focus on strategies such as developing hybrid models, preparing an input of less-noisy data, and using different model training methods to improve the simulation environment results. Ultimately, the proposed approach of creating simulation environments for DRL algorithms, utilizing SCADA data with a sufficient historical horizon to capture all system dynamics, is a promising way to improve process control in industrial applications.

\section{Acknowledgements}
The RecaP project has received funding from the European Union’s Horizon 2020 research and innovation programme under the Marie Skłodowska-Curie grant agreement No 956454. Disclaimer: this publication reflects only the author's view; the Research Executive Agency of the European Union is not responsible for any use that may be made of this information.

 \bibliographystyle{elsarticle-num} 
 \bibliography{cas-refs}





% \end{thebibliography}
\end{document}